\documentclass[pre,preprint,double-spaced,aps,apssymb,showkeys,showpacs]{revtex4}
\usepackage{amsmath,amssymb}
\usepackage{graphicx}% Include figure files
\usepackage{dcolumn}% Align table columns on decimal point
\usepackage{bm}% bold math
\usepackage{subfigure}
\usepackage{color}
\newcommand{\ber}{\begin{eqnarray}}
\newcommand{\eer}{\end{eqnarray}}
\newcommand{\bea}{\begin{equation}}
\newcommand{\eea}{\end{equation}}

\begin{document}

\title{Equilibrium of a Brownian particle with coordinate dependent diffusivity and damping: Generalized Boltzmann distribution}

\author{A. Bhattacharyay}
\email{a.bhattacharyay@iiserpune.ac.in}
\affiliation{Indian Institute of Science Education and Research, Pune, India}

\begin{abstract}
Fick's law for coordinate dependent diffusivity is derived. Corresponding diffusion current in the presence of coordinate dependent diffusivity is consistent with the form as given by Kramers-Moyal expansion. We have obtained the equilibrium solution of the corresponding Smoluchowski equation. The equilibrium distribution is a generalization of the Boltzmann distribution. This generalized Boltzmann distribution involves an effective potential which is a function of coordinate dependent diffusivity. We discuss various implications of the existence of this generalized Boltzmann distribution for equilibrium of systems with coordinate dependent diffusivity and damping.  
\end{abstract}
\pacs{05.40.Jc, 05.10,Gg, 05.70.-a}% PACS, the Physics and Astronomy
                             % Classification Scheme.
\keywords{Diffusion, Brownian motion, Coordinate dependent damping, Equilibrium, Fick's law.}%Use showkeys class option if keyword
\maketitle

\section{Introduction}
Coordinate dependence of diffusivity and damping \cite{bere} of Brownian particles in a confined liquid has been observed in many experiments \cite{fauc,lanc,volp}. In the context of protein folding, position dependent diffusion is supposed to play a major role \cite{best,humm}.  Position dependent diffusion is being taken into consideration even for hydrodynamics of optical systems \cite{yami,pauf}. In the context of equilibrium, coordinate dependent damping and diffusion are sources of long standing controversy \cite{lau,san1,san2,tupp,far1,far2,soko}. Brownian motion with coordinate dependent diffusion and damping, in general, is discussed in this paper. In such systems, homogeneity of space broken by space dependent damping and diffusion (boundary effects in small systems \cite{fauc}) can not be captured by a conservative force.  We show that the equilibrium probability distribution in such systems is a generalization of the Boltzmann distribution. The generalization is in the appearance of an effective potential which is a function of the coordinate dependent diffusivity. 
\par
This generalized distribution is normally overlooked as the equilibrium distribution of such systems in the conventional literature \cite{lau,san1,san2,tupp,far1,far2,soko}. To arrive at the Boltzmann distribution for such systems, in the standard literature \cite{lau,san1,san2,tupp,far1,far2,soko}, one effectively considers the diffusion current to be Fickian. Whereas, in the Smoluchowski dynamics the diffusion current comes out to be non-Fickian. In this paper, we try to resolve this controversy by giving an alternative derivation of the non-Fickian form of the diffusion current based on the definition of local diffusivity. This derivation assumes importance in view of the controversy over the form of diffusion current \cite{mark,lau} or effective consideration of the Fickian current to arrive at Boltzmann distribution \cite{san1,san2,tupp,far1,far2,soko}.  
\par
The paper is organized in the following way. We first present a short discussion of standard Brownian motion in the presence of constant diffusion and damping. Then we discuss the Kramers-Moyal expansion and coefficients to outline the standard derivation of the drift and diffusion coefficients in the It\^o/Stratonovich \cite{ito,stra} convention. After that, we show the derivation of non-Fickian diffusion current starting from the definition of the local diffusivity and obtain the generalized Boltzmann distribution as the equilibrium solution of the standard Smoluchowski equation. We conclude the paper with a discussion on the implications of existence of this generalized Boltzmann distribution. 
\section{Brownian Motion with constant diffusion and damping}
The Langevin dynamics of a Brownian particle (of centre of mass $x$) with a damping constant $\Gamma$, diffusivity D, in equilibrium at the minimum of a potential U($x$) at temperature T is
\bea
\frac{\partial x}{\partial t} = -\frac{1}{\Gamma}\frac{\partial \text U(x)}{\partial x} + \frac{\sqrt{{\text {2kT}}\Gamma}}{\Gamma}\eta(t).
\eea  
In this dynamics, k is the Boltzmann constant and $\eta(t)$ is a Gaussian white noise of zero mean. The part of the stochastic term under square root is present to ensure eventual convergence at large times to equilibrium characterized by the Boltzmann distribution. Considering diffusion alone in the absence of the confining potential U($x$), diffusivity of the system can be identified easily to follow the Stokes-Einstein relation ${\text D}=\frac{\text {kT}}{\Gamma}$ \cite{crank}. This identification practically fixes the strength of the stochastic term which is also extensively looked at from the perspective of fluctuation-dissipation theorem \cite{kubo}.  
\par
The Fokker-Planck equation (also called the Smoluchowski equation in the context of over-damped dynamics) for the dynamics of the probability density $\rho(x,t)$ of the Brownian particle is given by
\bea
\frac{\partial \rho(x,t)}{\partial t} = \frac{\partial}{\partial x}\left [\frac{\rho(x,t)}{\Gamma}\frac{\partial \text U(x)}{\partial x} + \text D\frac{\partial \rho(x,t)}{\partial x} \right ] = -\frac{\partial}{\partial x}j(x,t).
\eea
This is the equation of continuity for the conserved probability where $j(x,t)$ is probability current density. For a non-equilibrium stationary state the divergence of the probability current density has to vanish, whereas, for equilibrium $j(x,t)$ should be zero everywhere in space to maintain detailed balance. Since the space is inhomogeneous in the presence of a conservative force, presence of any probability current will produce entropy which cannot be there in equilibrium. This is where the local cancellation of the drift produced by conservative force $\text F(x)=-\frac{\partial\text U(x)}{\partial x}$ and diffusion component of the probability current is required in equilibrium. Equilibrium distribution as obtained by balancing the drift and diffusion currents is the Boltzmann distribution $\rho(x) = {\text N}e^{-\frac{\text U(x)}{\Gamma \text D}} = \text Ne^{-\frac{\text U(x)}{\text{kT}}}$ where N is a suitable normalization constant. 
\par
Keeping in mind the fundamental requirement of the local vanishing of the probability current in equilibrium, the next most important thing is the Fick's law that provides the structure of the diffusion current. Having the Fick's law in place, the equilibrium distribution is completely determined by detailed balance to be $\rho(x) = {\text N}e^{-\frac{\text U(x)}{\Gamma \text D}}$. The Stokes-Einstein relation provides additional structure and brings in the Boltzmann distribution through the introduction of the temperature. However, neither the Stokes-Einstein relation nor the resulting Boltzmann distribution is an essential requirement for equilibrium because detailed balance never explicitly needs these conditions.
\par
Existence of coordinate dependent damping and diffusion brings in additional parameters on top of conservative force that break the homogeneity of space even in the absence of a conservative force. But, the Boltzmann distribution does not at all reflect this reality. It completely disregards the presence of these additional symmetry breaking agents and hence cannot, in general, represent equilibrium distribution for systems with coordinate dependent damping and diffusivity. 
\section{Kramers-Moyal expansion and the Smoluchowski equation}
The Smoluchowski equation is the one obtained by truncating the Kramers-Moyal expansion at the term containing the second Kramers-Moyal coefficient and that is a very standard result. The dynamics of the probability density as obtained from the Kramers-Moyal expansion can be written as
\bea
\frac{\partial P(x,t)}{\partial t} = \sum_{n=1}^{\infty}{\left (-\frac{\partial}{\partial x}\right )^nD^{(n)}(x,t)P(x,t)},
\eea
where the Kramers-Moyal expansion coefficients are given by
\bea
D^{(n)}(x,t) = \frac{1}{n\,!}\lim_{\tau\to 0}\frac{1}{\tau}<[\xi(t+\tau)-x]^n>
\eea
with $\xi(t)=x$ and the angular brackets represent average over noise \cite{risk}. By Pawula's theorem, the transition probability remaining positive, the number of terms Eq.3 can have on the right hand side are either for $n=1,2$ or infinitely many. The Fokker-Planck equation is the one which contains terms with $n=1$ and $n=2$ with the relevant coefficients $D^{(1)}(x,t)$ and $D^{(2)}(x,t)$ identified as the drift and the diffusion coefficients respectively. One can easily infer, going by this formal theory, that the diffusion current density is $j(x)=-\frac{\partial D(x)P(x)}{\partial x}$ for stationary diffusivity and probability density. The $\tau \to 0$ limit in the co-efficients of the Kramers-Moyal expansion can be perceived to define local quantities. This is why the derivation is general and applies to situations with coordinate dependent diffusion. Let us have a closer look at the terms $D^{(1)}(x,t)$ and $D^{(2)}(x,t)$ following standard textbook \cite{risk}. 
\par
If the nonlinear Langevin dynamics is given by 
\bea
\dot{\xi} = h(\xi,t) + g(\xi,t)\eta(t),
\eea
where $\eta(t)$ is a Gaussian white noise of unit strength which ensures a Markov process, then
\bea
\xi(t+\tau) -x = \int_t^{t+\tau}{[h(\xi(t^\prime),t^\prime)+g(\xi(t^\prime)\eta(t^\prime)]dt^\prime}.
\eea
For a small $\tau$,
one does a Taylor expansion such that
\ber\nonumber
h(\xi(t^\prime),t^\prime) = h(x,t^\prime) + \left [\frac{\partial h(\xi(t^\prime),t^\prime)}{\partial{\xi(t^\prime)}}\right ]_{\xi(t^\prime)=x}(\xi(t^\prime)-x) + ...\\
g(\xi(t^\prime),t^\prime) = g(x,t^\prime) + \left [\frac{\partial g(\xi(t^\prime),t^\prime)}{\partial{\xi(t^\prime)}}\right ]_{\xi(t^\prime)=x}(\xi(t^\prime)-x) + ...
\eer
Inserting Eq.7 in Eq.6 and iterating for $(\xi(t+\tau) -x)$, to the leading order in $\tau$ while taking a noise average one gets
\bea
<\xi(t+\tau) -x> = \int_t^{t+\tau}{h(x,t^\prime)dt^\prime} + \frac{1}{2}\int_t^{t+\tau}{\frac{\partial g(x,t^\prime)}{\partial x}g(x,t^\prime)dt^\prime}.
\eea
Taking the $\tau \to 0$ limit results in 
\bea
D^{(1)}(x,t) = h(x,t) + \frac{1}{2}\frac{\partial g(x,t)}{\partial x}g(x,t),
\eea
where, at this limit, all the sub-leading order terms vanish being $O(\tau)$ or even smaller.
An important point to note here is that (refer to chapter-3 of \cite{risk} for details), in arriving at the above expression for the $D^{(1)}(x,t)$, the $\delta$-function arising from noise average has been treated such that $\delta(t)=1/\epsilon$ for $-\epsilon/2<t<\epsilon/2$ and $\delta(t)=0$ elsewhere followed by the limit $\epsilon \to 0$. This way of treating the $\delta$-function is equivalent to considering the Stratonovich convention whereas adoption of It\^o convention would have resulted in $D^{(1)}(x,t) = h(x,t)$ (refer to chapter-3 of \cite{risk} for details). 
\par
The second part in the expression of the drift coefficient $\frac{1}{2}\frac{\partial g(x,t)}{\partial x}g(x,t)$ is in general referred to as the spurious current because it is an artefact of the convention used for stochastic integration involving a discontinuous function $\eta(t)$. In the present case, it is an artefact of the Stratonovich convention. On adoption of different conventions this term can be different, however, that should not matter much because it has in the end to be neglected. 
\par
However, the diffusion coefficient $D^{(2)}(x,t)$ turns out to be convention independent in the sense that there is no spurious contribution to $D^{(2)}(x,t)$ in any convention. This fact is clearly shown by a general treatment of the It\^o and Stratonovich conventions using Stieltjes integral and Wiener process representation of the noise in ref.\cite{risk}. Following the similar procedure as shown for the computation of $D^{(1)}(x,t)$, one gets the expression for $D^{(2)}(x,t)$ as
\bea
D^{(2)}(x,t) = \frac{1}{2}\lim_{\tau \to 0}\frac{1}{\tau}\int_t^{t+\tau}{g(x,t^\prime)\int_t^{t+\tau}g(x,t^{\prime\prime})\delta(t^\prime - t^{\prime\prime})dt^\prime dt^{\prime\prime}} = \frac{g^2(x,t)}{2},
\eea
which is the same for It\^o and Stratonovich conventions. The diffusion term, in the ensuing Smoluchowski equation, does not depend on convention. For any Markov process (linear or non-linear), the Smoluchowski equation can be arrived at following this formal procedure as described in the textbook \cite{risk}. 
\par
Note that, as Eq.7 is put into Eq.6, in the third and even higher order Kramers-Moyal coefficients, all the integrals appearing will result in terms at least $O(\tau^2)$ or smaller. As a result, at the $\tau\to 0$ limit, all the coefficients higher than second order will be zero because those are at least linear in $\tau$. This is exactly how the consistency of the structure of the Fokker-Planck equation with the Pawula's theorem gets established in the derivation following the Kramers-Moyal expansion.
\par 
In the following section we have a look at Fick's law for systems with coordinate dependent diffusivity from another perspective and then obtain the generalized Boltzmann distribution as the equilibrium solution of the Smoluchowski equation.

\section{Modified Fick's law and generalized Boltzmann distribution}
The consideration of non-Fickian diffusion is not something new in the literature, but, to the knowledge of the present author, has never been seriously considered in the context of true thermodynamic equilibrium. Any diffusion which is non-Fickian is, in general, considered to bear in it non-equilibrium signature of the system \cite{crank}. We show here a simple derivation of the modified Fick's law for coordinate dependent diffusivity starting from the definition of diffusivity (first principle) and that modification will turn out to be exact at the local limit. We are concerned with finding the diffusive part of the probability current density. So, any density referred to in the following has to be understood as probability density.   

Global diffusivity is defined as
\bea
\text D=\lim_{t\to \infty}\frac{1}{2t}<[x(t)-x(0)]^2>,
\eea   
where $t$ represents time, $x(0)$ and $x(t)$ are the positions of the particle at times $t=0,t$ respectively. The angular brackets indicate an average over many realizations of stochastic noise. Existence of the long time limit makes diffusivity a global quantity. To make it local one has to drop the limit and the time scale will be determined by the length scale over which the local diffusivity is getting defined. The discussion on the diffusion coefficient in the previous section implies that this limit should actually be $t \to 0$ for local diffusivity. Keeping that in mind, by dropping the limit over time, the local diffusivity can be identified as $D(x(0))=\frac{1}{2t}<[x(t)-x(0)]^2>$ where a translation of the initial position while doing the noise average is not allowed because the space is inhomogeneous, however, homogeneity of time can be assumed (since we are concerned here with locality over space) to consider the diffusivity to be time independent for the sake of simplicity.

Consider that this local diffusivity, which dimensionally is a product of a length and velocity, exists. Refer to schematic Fig.1 (for clarity) where the smallest length scale over which diffusivity and density are considered to change is $\delta x$ and not on scales smaller than this. We are interested in finding out the diffusion current $j$ in the middle of coordinates $x$ and $x+\delta x$ (i.e. at $x+\frac{\delta x}{2}$) due to the difference in probability of finding the Brownian particle between the regions $R_l$ lying between $x-\frac{\delta x}{2}$ to $x+\frac{\delta x}{2}$ and $R_r$ within $x+\frac{\delta x}{2}$ to $x+\frac{3\delta x}{2}$. A suitable uniform unit cross-sectional area along x-axis is under consideration. Since the diffusivity and density do not change appreciably (by definition) over the length scale $\delta x$, consider the density and diffusivity at the middle of the regions $R_l$ and $R_r$ to represent the quantities within these domains. 

\begin{figure}[h]
  \includegraphics[width=8cm]{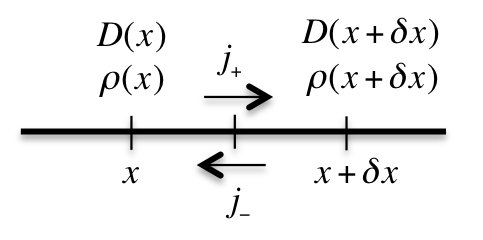}
  \caption{\small {Schematic diagram for local version of Fick's law.}}
\label{fig:Local diffusivity}
\end{figure}

Keeping in mind the dimension of the diffusivity, one can write the current $j_+$ at $x+\frac{\delta x}{2}$ due to the region $R_l$ as $j_+=\frac{\text D(x)}{\delta x}\rho(x)$ and that at the same point the current density in the negative direction due to $R_r$ as $j_-=\frac{\text D(x+\delta x)}{\delta x}\rho(x+\delta x)$. The factor of half in the definition of diffusivity nicely takes care of the left-right symmetry in the regions $R_l$ and $R_r$ as considered practically homogeneous. Therefore, the total diffusion probability current density at the middle is 
\bea
j=j_+-j_-=-\frac{\partial\text D(x)\rho(x)}{\partial x} + O(\delta x)
\eea 
where a Taylor expansion upto the first order is explicitly considered. To this order of accuracy the above equation gives the modified Fick's law for coordinate dependent diffusivity based only on the definition of diffusivity and its existence. The correction terms to the modified Fick's law (as appear in the above equation) are $O(\delta x)$ or even smaller and will vanish on taking the local limit which must be the case for the resulting Fokker-Planck equation to be consistent with Pawula's theorem. Therefore,
\bea
\lim_{\delta x \to 0}j = -\frac{\partial\text D(x)\rho(x)}{\partial x}
\eea
is the exact form of the Fick's law for coordinate dependent diffusivity.
\par
Having found the modification in the Fick's law in the presence of coordinate dependent damping let us consider the dynamics of probability which will be a simple generalization of Eq.2 as

\bea
\frac{\partial \rho(x,t)}{\partial t} = \frac{\partial}{\partial x}\left [\frac{\rho(x,t)}{\Gamma(x)}\frac{\partial \text U(x)}{\partial x} + \frac{\partial\text D(x) \rho(x,t)}{\partial x} \right ] = -\frac{\partial}{\partial x}j(x,t).
\eea
This exactly is the Smoluchowski equation that one will arrive at using the standard procedure of Kramers-Moyal expansion. 
\par
Setting $j(x)=0$ for equilibrium and identifying the conservative force $\text F(x) = -\frac{\partial \text U(x)}{\partial x}$, the equilibrium probability distribution can readily be written as
\bea
\rho(x) = \frac{\text N}{\text D(x)}\exp{\int_{-\infty}^x \frac{\text F(x^\prime)}{\text D(x^\prime)\Gamma(x^\prime)}dx^\prime}.
\eea
This is the most general equilibrium probability distribution with a normalization constant N in an inhomogeneous space where the diffusivity and damping are coordinate dependent. If one considers the local validity of Stokes-Einstein relation one gets the equilibrium distribution involving the temperature of the bath the system has equilibrated with as
\bea
\rho(x) = {\text N^\prime}{\Gamma(x)}\exp{-\text U(x)/\text {kT}},
\eea
where N$^\prime$ is another normalization constant. The distribution differs from the Boltzmann distribution in having a coordinate dependent amplitude. However, one should also keep in mind that the Stokes-Einstein relation comes from fluctuation-dissipation relation and the Boltzmann distribution. Fluctuation-dissipation relation being a result of causality should always hold. However, its form involving an average using the Boltzmann distribution may not hold when the Boltzmann distribution itself gets altered or modified.
\par
Let us, therefore, speculate about another form where the diffusivity is a constant despite having the damping coordinate dependent, for example if $\text D = \text{kT}/<\Gamma(x)>$ the distribution is
\bea
\rho(x) = {\text N^{\prime\prime}}\exp{\frac{<\Gamma(x)>}{\text{kT}}\int_{-\infty}^x \frac{\text F(x^\prime)}{\Gamma(x^\prime)}dx^\prime},
\eea
which is again a modified Boltzmann distribution. This distribution has a coordinate dependent effective temperature $T_l = T\Gamma(x)/<\Gamma(x)>$ even in equilibrium and has already been considered by the present author as an alternative equilibrium scenario in \cite{ari1,ari2,ari3}. This is a nice case where the corresponding Langevin dynamics can easily be mapped to that in homogeneous space and standard equilibrium theory of Brownian motion can be brought to bear. This obviously is an exotic equilibrium solution with coordinate dependent effective temperature where equipartition of energy holds on an average over the entire phase space.
\par
By writing the probability distribution (Eq.15) in a Boltzmann distribution form - {\it generalized Boltzmann distribution for equilibrium} - one can identify an effective potential as $V(x)=\ln{D(x)}-\int_{-\infty}^x{dx^\prime F(x^\prime)/D(x^\prime)\Gamma(x^\prime)}$ up to a multiplicative constant. This effective potential is not the same as $U(x)=-\int_{-\infty}^x{dx^\prime F(x^\prime)}$ which is used in the conventional Boltzmann distribution. Using this effective potential one can write down a Langevin dynamics with additive Gaussian white noise in the place of a multiplicative noise problem. After all, the purpose of the Langevin dynamics is to simulate equilibrium fluctuations. Therefore, once the distribution is known, one can take the simplest path to sample equilibrium fluctuations.

\section{conclusion}
The Smoluchowski equation as derived using standard methods like Kramers-Moyal expansion presents the dynamics of probability of systems with coordinate dependent diffusivity and damping. The modification of Fick's law and diffusion current thereof is justified in the presence of coordinate dependent diffusivity. The resulting equilibrium distribution is a generalization of the Boltzmann distribution. This generalized distribution is a function of an effective potential whose structure may change depending upon the relationship (akin to Stokes-Einstein relation in the homogeneous case) between the diffusivity and damping in varied circumstances. One of such form is shown in Eq.17 manifests a local temperature in equilibrium. The local temperature is identified as $T_l = T\Gamma(x)/<\Gamma(x)>$. In this case, it has been shown in ref.\cite{ari2,ari3} that the equipartition of energy holds on an average over the whole phase space. Other such scenarios might also be possible and equilibrium distribution may show variety within the scope of the generalized Boltzmann distribution.

One of the long standing unsolved problems in classical mesoscopic systems is Levinthal's paradox in protein folding. Possibly, a modification of the equilibrium distribution holds the clue as to how a path is cut over a rugged energy landscape to make a protein quickly find its global minimum corresponding to the native fold. People have started identifying the role of space dependent damping and diffusion in this context \cite{best,humm}, but, a possible modification in the Boltzmann distribution is not yet considered. There should not be much doubt in the fact that the equilibrium fluctuations play a major role in protein folding, because, it actually is not an active process. However, the equilibrium distribution used mostly in simulations is the Boltzmann distribution.

There is no reason for the present results to only apply to mesoscopic systems. The general derivation does not take into account the size of the system. Therefore, other broad area of physics which can receive considerable attention in light of the modification of equilibrium results pertaining to Brownian motion probably is critical dynamics. The basic assumption in the treatment of critical dynamics of slow modes, where the drift term is derived from a free energy functional, is that the slow modes are in near-equilibrium state with the bath created by the faster modes. The consideration of Boltzmann distribution to do near equilibrium averages is quite standard in critical dynamics. So, any modification in the distribution function can affect results here. Modification of Boltzmann weight can affect quantum systems with such inhomogeneities where density matrix would involve the generalized Boltzmann weight. Experimental verification of this generalized Boltzmann distribution is, therefore, of immense importance and should possibly be done by carefully watching Brownian motion near a wall as has been described in ref.\cite{ari3}.

\end{document}